# VO$_2$ under hydrostatic pressure: Isostructural phase transition close to a critical end-point

P. Bouvier[1*], L. Bussmann[1], D. Machon[2,3,4], I. Breslavetz[5], G. Garbarino[6], P. Strobel[1], V. Dmitriev[1]

[1] *Université Grenoble Alpes, Institut Néel CNRS, 25 Rue des Martyrs, 38042, Grenoble, France.*
[2] *Laboratoire Nanotechnologies et Nanosystèmes (LN2) - CNRS UMI-3463 Institut Interdisciplinaire d'Innovation Technologique (3IT), Université de Sherbrooke, 3000 Boulevard Université, Sherbrooke, J1K OA5 Québec, Canada*
[3] *Institut Interdisciplinaire d'Innovation Technologique (3IT), Université de Sherbrooke, 3000 Boulevard Université, Sherbrooke, J1K OA5 Québec, Canada*
[4] *Université de Lyon, INSA Lyon, CNRS, Ecole Centrale de Lyon, Université Claude Bernard Lyon 1, CPE Lyon, INL, UMR5270, 69621 Villeurbanne, France*
[5] *LNCMI, UPR 3228, CNRS, EMFL, Université Grenoble Alpes, 38000 Grenoble, France*
[6] *European Synchrotron Radiation Facility, BP220, 38043 Grenoble Cedex, France.*

* Corresponding author. *E-mail address:* pierre.bouvier@neel.cnrs.fr (P. Bouvier).

**Abstract**

The high-pressure behavior of monoclinic VO$_2$ is revisited by a combination of Raman spectroscopy and X-ray diffraction on a single crystal under hydrostatic conditions at room temperature. A soft mode is observed up to P$_c$= 13.9(1) GPa. At this pressure, an isostructural phase transition between two monoclinic phases M$_1$ and M$_1$' hinders this instability. The features of this transformation (no apparent volume jump) indicate that the compression at ambient temperature passes close to a critical point. An analysis based on the Landau theory of phase transitions gives a complete description of the P-T phase diagram. The M$_1$' is characterized by spontaneous displacements of the oxygen sub-lattice without any strong modification of the VV dimers distances nor the twist angle of vanadium chains. The spontaneous displacements of oxygen and the spontaneous deformations of the ($b_{M1}$, $c_{M1}$) plane follow the same quadratic dependence with pressure and scales with spontaneous shifts of the Raman phonons located at 225, 260 and 310 cm$^{-1}$. Pressure-induced shifts of the Raman peaks allows for new assignment of several Raman modes. In particular, the A$_{g(1)}$+B$_{g(1)}$ modes at 145 cm$^{-1}$ are identified as the vanadium displacive phonons. A second transformation in the metallic phase X, which is found triclinic ($P\bar{1}$) is observed starting at 32 GPa, with a wide coexistence region (up to 42 GPa). Upon decompression, phase X transforms, between 20 GPa and 3 GPa, to another phase that is neither the M$_1$' nor M$_1$ phase. The structural transitions identified under pressure match with all the previously reported electronic modifications confirming that lattice and electronic degrees of freedom are closely coupled in this correlated material.

## 1. Introduction

VO$_2$ is a well-known prototypical electron-correlated material, showing a Metal-to-Insulator Transition (MIT) at ambient pressure and moderate temperature T= 340K [1] accompanied with a structural phase transition. Despite VO$_2$ is already used in a variety of technological applications, such as infrared detection, thermochromics, transistors or microactuators (see the reviews [2,3,4]), the microscopic mechanism of the MIT is still an open fundamental question and a challenge for finding accurate functionals for theoretical DFT calculations [5,6]. Two mechanisms have been proposed and are still debated in many experimental and theoretical studies: the Peierls lattice distortion model and the Mott orbital electron model (or a mixture of both mechanisms) [7,8,9,10,11,12,13,14,15,16,17,18,19,20,21,22,23,24,25,26,27,28,29,30,31,32,33].

The associated structural transition from the metallic rutile structure ($P4_2/mnm$, n°136, Z=2 [34]) to the low-temperature insulating monoclinic ($P2_1/c$, n°14, Z=4 [35]), named $M_1$, was explained by the phonon condensation at the R-point of the rutile Brillouin zone with vanadium displacements as the order-parameter (OP) [36,37,10,11,38,39]. Thus, the metallic rutile structure is made of two vanadium chains with equal VV distances whereas the insulating monoclinic phase is characterized by two zigzagging chains with VV dimers. The thermodynamics of this displacive Peierls mechanism and the stability limits of the different phases were described in the framework of a Landau-type phenomenological model with a reduced two-dimensional component OP and free-energy expanded to six-degree, and eventually coupled with the strain [40,41,42,43,44,45,46,32,47]. This phenomenological description predicts the possibility of stabilizing other phases, such as a monoclinic $C2/m$ (n°12) phase, named $M_2$, and an intermediate triclinic phase $P\bar{1}$ (n°2), named T (or $M_3$) [10,38,39]. These $M_2$ and T structures were observed in $VO_2$ doped with cation of lower oxidation states [48,49,50,51,52,45] or under specific uniaxial stress [53,34,54,55,41,42,43,56,57,58,59,60,61,46,62]. The MIT was found to be remarkably affected by mechanical stresses [63,64,65] and a triple point between $M_1$, $M_2$ and rutile phases was observed at 340 K at zero strain [40,58].

Applying pressure is a relevant way to modify the stability between structural or electronic degrees of freedom. Thus, in $M_1$ phase of $VO_2$, spectral discontinuities in both the mid-infrared optical conductivity and in the behavior of two Raman-active phonons located at 190 and 225 cm$^{-1}$ [66,67,68,69], observed at 10 GPa under quasi-hydrostatic pressure, were interpreted as vanadium dimers rearrangement [66]. Electrical discontinuity was also reported at 10-13 GPa [70,71]. Synchrotron X-ray diffraction studies of pure $VO_2$ powders [72,71,69,73] or nanoparticles [74] have shown that the $M_1$ phase transforms, above 11-13 GPa, to an isostructural phase (same space group $P2_1/c$ n°14, Z=4), named $M_1$'. Since there is no apparent change in the crystal symmetry, the transition pressure is defined by a discontinuity in the compression behavior of the ($b_{M1}$, $c_{M1}$) monoclinic plane [72,71,69,74]. Contrary to early studies, Bai at al. proposed that the discontinuities measured at the $M_1$-$M_1$' transition in the pressure dependence of the Raman modes located at 190, 225 and 320 cm$^{-1}$ are not associated to any V-chains rearrangement [71]. The persistence of the VV dimerization up to 22 GPa and a VV pair twist angle remaining close to 3° was then confirmed by atomic pair distribution function analysis [75]. Density functional theory calculations suggested that the $M_1$-$M_1$' transition is induced by an unstable Γ-point phonon that is related to the rotation of the oxygen octahedra along the monoclinic $a_{M1}$ axis (or the parent rutile $c_R$ axis) [76]. In their calculations, the pressure-induced reduction of the band gap and metallization is accounted by clockwise rotations (phase $M_1$'') that progressively reduce the dimerization and zigzags of the vanadium chains [76].

At higher pressure, a second phase transition to a metallic phase X was detected (between 28-50 GPa). The slope change [69] or splitting [71] of the Raman mode at 225 cm$^{-1}$, observed above 27.8 GPa, was assigned to phase X. Afterwards, Balédent et al. observed this splitting at 19 GPa and proposed a new insulating $M_3$ phase, different from the metallic phase X [73]. Different structures have been proposed for phase X such as a monoclinic baddeleyite-type ($P2_1/c$, n°14) with Z=8 [71,69] or with Z=4, named Mx [74,77] in which the vanadium coordination number increases from six to seven. Xie et al. proposed that a different seven-coordinated orthorhombic structure ($Pmn2_1$, n°31, Z=2) coexists with the low-pressure $M_1$ between 29 and 79 GPa [78]. A different monoclinic space group ($Pn$, n°7) was inferred using spin-polarized *ab initio* structure search [73]. Under decreasing pressure, another monoclinic baddeleyite-type polymorph, named Mx', was reported following a high-pressure treatment of the $M_1$ phase up to 63 GPa [74,77]. Additional pressure measurements, at 383 K [71], or on W-doped $VO_2$ [79], have shown that rutile phase transforms at 13.3 GPa to an orthorhombic $CaCl_2$-type structure ($Pnnm$, n°58, Z=4) and coexists with metallic phase X between 32 and 64 GPa. The pressure-temperature phase diagram of $VO_2$ was built using Raman, optical reflectance and electrical transport characterizations [80].

Therefore, although many experimental and theoretical calculations were published concluding on the presence of several different $M_1'$, $M_1''$, $M_3$, phase X, Mx and Mx' structures under increasing pressure, no agreement has yet been reached on the phase sequence under high pressure and on the associated mechanisms. One of the reasons lies in the experimental limitation due to the form of the sample (powder, nanobeams) and quasi-hydrostatic conditions that can play a significant role. The aim of this study is to present new results obtained by Raman and X-ray diffraction analysis on a high-quality $VO_2$ single crystal compressed under hydrostatic conditions using Helium as the pressure-transmitting medium. In a first section, X-ray diffraction data obtained during compression will be displayed. These make it possible to clarify the phase transition sequence and the microscopic mechanism involved. Then, Raman spectroscopy measurements will be presented with new insights in terms of assignment and pressure-induced behaviour. A phenomenological analysis will be proposed to describe the experimental P-T phase diagram of $VO_2$. In the last section, the obtained results are combined to correlate the behaviours of the Raman modes with the strains and microscopic characteristics of the compound. This will be of interest to characterize phases in thin films of (doped) $VO_2$ and the nature and amplitude of the strains.

2. **Experimental**

High-quality crystals with natural faces of stoichiometric $VO_2$ crystals were produced by chemical vapour transport, using $TeCl_4$ transport agent and following the procedure described in Ref. [81].

High-pressure experiments were performed using a membrane driven diamond anvil cell (DAC) with 250/300 µm bevelled diamond culets. A pressure chamber of 160 µm in diameter and 40 µm in thickness was drilled in a stainless-steel gasket. Helium, loaded at 1.4 kbar, was used as the pressure transmitting medium to ensure high hydrostatic pressure conditions up to 42 GPa, the highest pressure reached in this study. During the Raman experiment, the pressure was measured using the $R_1$-line emission of a ruby ball placed close to the sample using Holzapfel equation of state [82]. The ruby signal is measured before and after each measurement in order to control the pressure drift during long acquisitions. The recorded pressure is set at the average of these two pressure values and the uncertainty is set as the half of the difference between these two values. The homogeneity of the pressure in the DAC was followed from both the width and the splitting between the $R_1$ and $R_2$ ruby lines [83,84]. During the X-ray diffraction experiment, the pressure was measured using the equation of state of pure copper powder [85] placed close to the crystal. The copper X-ray diffraction images were integrated with the Dioptas software [86].

Three experiments on two different single crystals were done. During the first one, we recorded only Raman on a crystal of 40x30 µm in size and 10 µm in thickness up to 42 GPa and back to room pressure. We have reproduced this Raman experiment on a smaller crystal of 15x18 µm and 10 µm thickness up to 25 GPa and back to room pressure. During this second experiment, we have chosen not to exceed 25 GPa in order to avoid forming the high-pressure metallic phase. A third experiment up to 35 GPa, using X-ray diffraction was done with the second crystal that had already experienced pressure during the second Raman experiment.

The ruby and Raman measurements were made at room temperature using a 514.4 nm laser (Cobolt Fandango) and a 750 mm spectrometer (SP2750, Acton Research) with a 2400 grooves/mm grating (blazed at 500 nm), equipped with a cooled CCD camera (PyLoN, Princeton), and a 50 µm entrance slit size that provides a resolution of 0.70 cm$^{-1}$ (0.019 nm). A set of Bragg filters (BNF-Optigrate) were used in order to reject the excitation line. The spectra were recorded in backscattering geometry with a 50X objective (Nikon) to focus the incident laser beam and collect the scattered light from inside the DAC through the diamond anvil. The spectrometer was calibrated in wavenumber using the lines of a Ne-Ar lamp. The incident laser power was fixed at 0.5 mW (measured before the DAC) in order to avoid any laser heating of the sample that could induce the $M_1$-Rutile transition at 340 K. The Raman spectra covering a 25–900 cm$^{-1}$ spectral range were recorded using two monochromator positions with a maximum of 300 s acquisition time averaged over two to four acquisitions. In the 25-150 cm$^{-1}$ range, we have subtracted the contribution of $N_2/O_2$ rotations lines. Spectral parameters (position and full-width at half maximum FWHM) were obtained from the decomposition of each spectrum with several Lorentzian peaks using Fityk software (version 1.3.1) [87].

Single crystal X-ray diffraction (XRD) experiment was done at ID15B beamline (ESRF Grenoble) with a monochromatic wavelength λ=0.41020 Å and a 2x4µm focused beam. Diffraction images were collected during the continuous rotation of the DAC around the vertical ω axis in a range ±32°, with an angular step of Δω=0.5° and an exposure time of 0.5 s/frame. The CrysAlis$^{Pro}$ software package [88] was used for the analysis of the single-crystal XRD data (indexing, data integration, frame scaling, and absorption correction). A single crystal of Vanadinite [$Pb_5(VO_4)_3Cl$, *Pbca* space group, *a* = 8.8117(2) Å, *b* = 5.18320(10) Å, and *c* = 18.2391(3) Å] was used to calibrate the instrumental model in the CrysAlis$^{Pro}$ software, i.e., the sample-to-detector distance, detector's origin, offsets of the goniometer angles, and rotation of both the X-ray beam and detector around the instrument axis. Using the Jana2006 software package, the structure was solved with the ShelXT structure solution program [89]. Crystal structure visualization was made with the VESTA software [90]. The equation of state was obtained by fitting the pressure–volume data using a third order Birch-Murnaghan (BM EoS). Le Bail profile analyses of the pattern measured at 35 GPa have been carried out using the FULLPROF software [91]. Cell parameters and overall thermal factor are refined. The background was first removed with a spline interpolation and then refined as a linear function. The peak shape was described with pseudo-Voigt function. The profile parameters *u,v,w* and the mixing parameter of the pseudo-Voigt function were kept fixed for the final refinement.

### 3. Results

#### *3.1. Single crystal X-ray diffraction under high-pressure*

The single crystal diffraction measured in the restricted geometry of the DAC allows to index 180 peaks (~30 % of the total reciprocal lattice) in a monoclinic reduced niggly-cell with $a_{M1}$= 5.3548(6) Å, $b_{M1}$= 4.5253(2) Å, $c_{M1}$= 5.3817(3) Å, $β_{M1}$= 115.224(9)° and volume $V_{M1}$=117.974(15) Å$^3$ with space-group $P2_1/c$ (n°14, Z=4, cell choice 1). Notice that this reduced cell is identical to the $P2_1/n$ (n°14, Z=4, cell choice 2) monoclinic cell with $a_{M1}$= 5.7510(8) Å, $b_{M1}$= 4.5253(17) Å, $c_{M1}$= 5.3548(6) Å, $β_{M1}$= 122.16(2)° in agreement with the lattice parameters reported in the ICSD [35] for phase $M_1$. The reciprocal maps attest to the absence of multi domains in the crystal measured under pressure (see Figure S1 in the supplementary information). Unfortunately, the orientation of the crystal in the DAC was not favourable to access to the [0k0] direction in the (hk0) plane and to confirm the presence of the $2_1$-screw axis along the b axis. However, the specific extinctions (h0l) with h+l=2n and (h00) with h=2n due to the presence of a mirror *n* perpendicular to the b axis are observed. The crystallographic extinctions are not modified up to 34 GPa (Figure S2) which discard any structural transition to $P\bar{1}$ (n°2), $P2_1$ (n°4), or *Pc* (n°7) subgroups of the $P2_1/c$ space group. The diffraction intensities are refined in the $M_1$ phase (see the refinement parameters at 0.3 GPa in Table S1). The crystallographic parameters (unit cell parameters, volume and atomic positions) up to 34 GPa are given in Table S2.

Figures 1(a-d) display the monoclinic unit cell parameters evolution with increasing pressure. As observed previously [72,71,69,74], the $a_{M1}$ lattice parameter decreases without any detectable discontinuity between 0 and 34 GPa whereas, above 13-14 GPa, a discontinuity is observed in the ($b_{M1}$, $c_{M1}$) monoclinic plane, i.e., the $b_{M1}$ softens while the $c_{M1}$ hardens simultaneously. A discontinuity is also observed in the pressure behavior of the beta angle at 14 GPa (see Fig 1c). The non-linear pressure dependence of cell parameters are reproduced by a third order Birch-Murnaghan-like equation of state (BM EoS) with $a_{M1}°$= 5.7506(7) Å , K°= 545(5) GPa and K'=4.9(3) between 0 and 34 GPa and by three second order BM-like EoS with $b_{M1}°$= 4.5259(7) Å , K°= 630(9) GPa, $c_{M1}°$= 5.3521(12) Å , K°= 820(23) GPa and β°= 122.16(1)° , K°= 12672(682) GPa between 0 and 13 GPa. Below 14 GPa, the monoclinic $a_{M1}$ cell parameter is more compressible than the $b_{M1}$ and $c_{M1}$ parameters and the beta angle is remarkably stiff. Using the EoS of the low pressure M$_1$ and extrapolating them above 14 GPa, the elastic spontaneous deformations e$_{11}$, e$_{22}$, e$_{33}$, e$_{12}$, e$_{13}$, e$_{23}$ and the e$_{total}$=√(∑e$_{ij}^2$) are calculated in the high-pressure monoclinic phase. The e$_{11}$, e$_{12}$, e$_{13}$, e$_{23}$ stay at values close to zero whereas the e$_{22}$, e$_{33}$ and e$_{total}$ increase as the square root of (P-P$_c$) as shown in Figure 2. Maximum values of e$_{22}$=-1.5%, e$_{33}$=+2.5% and e$_{total}$=+2.9% are reached at 34 GPa.

The pressure dependence of the volume shown in Figure 3(a), did not show any obvious discontinuity in the whole pressure range. The volume variation was first fitted with one unique third order BM EoS with V$_0$=117.97(4) Å$^3$, K°= 214(2) GPa and K'= 2.5(1) between 0 and 34 GPa. However, the value of K' less than 4 and the discontinuity at 13-14 GPa in the F-f plot reveal the structural transition (see insert in figure 3(a) using V$_0$=117.97 A$^3$). Thus, the EoS of M$_1$ phase are V$_0$=118.00(4) A$^3$, K°= 194(7) GPa and K'= 7(1) between 0 and 14 GPa and V$_0$=119.6(6) A$^3$, K°= 162(17) GPa and K'= 4.6(8) between 14 to 34 GPa. The EoS of M$_1$ and M$_1$' phases agree with Ref. [71]. The K° values are 15% lower than those measured on nanoparticles [74]. The distance between the two vanadium atoms of VV dimers along the monoclinic chain shows a regular decrease with pressure from 2.62 Å to 2.47 Å at 34 GPa (see Figure 3(b)) and is fitted by a third order BM-like EoS with d$_{VV}$°= 2.6199(8) Å, K°= 564(14) GPa and K'=2.3(8). A maximum contraction of 5.7% is measured at 34 GPa. As shown in Figure 3(c) the VO$_6$ polyhedra reduce their volume without any apparent discontinuity at 14 GPa and can be reproduced by a third order BM EoS with V°$_{octa}$= 9.542(6) Å$^3$, K°= 173(5) GPa and K'=12.2(7). A maximum contraction of 10% is measured at 34 GPa. The individual VO distances inside an octahedron, reported in Figure S3, show a regular decrease with the tendency for the VO$_6$ polyhedron to become more symmetric.

The relative variation of the atomic fractional parameters with pressure obtained from refining the single crystal diffraction intensities indexed in space-group $P2_1/n$ (n°14, Z=4, cell choice 2) are reported in Figure 4. Vanadium and oxygen atoms are in general position (site 4e). The vanadium coordinate along $b_{M1}$ increases continuously by 0.4(1)% at 34GPa. They decrease by 0.3(1)% at 14 GPa in the ($a_{M1}$, $c_{M1}$) plane and remain constant above. The two oxygen fractional positions almost do not change in the pressure range 0-14 GPa, but they display a clear deviation above 14 GPa that is one order of magnitude larger than that of the vanadium displacements. In figure 4(b), we report the spontaneous displacements of both oxygens atoms measured along the three crystallographic directions after subtracting the displacements extrapolated from the behavior below 14 GPa. Notice that both oxygen atoms display opposite spontaneous displacements of the exact same amplitude along $a_{M1}$ (former c$_R$ axis in the rutile phase) and $c_{M1}$ directions while they move in the same direction along $b_{M1}$ (former a$_R$ axis in the rutile phase). The oxygen spontaneous displacements follow a square root dependence with P-P$_c$ with fixed P$_c$=13.9 GPa as plotted with plain lines in figure 4(b).

At 35 GPa, the previous well resolved single crystal diffraction pattern disappeared suddenly. The crystal is damaged which indicates a first order transition. Some crystallographic axes are still observed;

however, Bragg peaks are spread in the azimuthal direction (see insert in Figure 5). Different structural models (including baddeleyite-type phase X from ref [71], Mx from ref. [74,77], or orthorhombic phase from ref [78]) were tested but none of them can reproduce the X-ray diffraction pattern. The pattern was indexed with a triclinic ($P\bar{1}$) cell with $a$=9.075(3) A, $b$=4.412(2) A, $c$=4.996(3) A, $α$=87.84(4)°, $β$=94.52(4)°, $γ$=92.67(4)° and V=199.05(19) A$^3$ with Bragg R factor of 0.4% as reported in Figure 5. The unit cell contains height $VO_2$ formula unit. A volume jump of ΔV/V=-3.3(1) % is measured at the transition. The high-pressure phase X is different from the structural model reported for the triclinic phase in $VO_2$ doped with cation of lower oxidation states or under uniaxial stress. If it was the case, we would expect a second order continuous transition that is not observed. Attempts were made to refine the structure starting from a baddeleyite-type model but the statistics in azimuthal direction was not good and the intensity was too low to obtain a reliable refinement.

### *3.2. Single crystal Raman spectra under high-pressure*

The Raman spectrum measured on a $VO_2$ single crystal is identical to previously published spectra for the $M_1$ phase [92,93,94,95,96,97]. Eighteen Raman active modes (9$A_g$+9$B_g$) are expected and almost all of them were identified at 83 K on a naturally oriented single crystal [93,94] (Table 1). Figure 6 displays a zoom on the low wavenumber part of the Raman spectra (70-340 cm$^{-1}$) to highlight the softening/hardening of the low-laying 145 cm$^{-1}$ weak mode observed under pressure. The stokes and anti-stokes spectra measured at 21 GPa (see figure S5) confirm that this mode is a phonon and not a fluorescent artefact. The entire Raman spectra measured up to 25 GPa, are reported in figure S4. In this work the Raman modes are labelled as $A_{g(1)}$ to $A_{g(9)}$, and $B_{g(1)}$ to $B_{g(9)}$ in Figure 6 and in Figure S4.

Figure 7 presents the pressure dependence of spectral parameters obtained from the decomposition of Raman spectra with Lorentzian functions. In the past studies, the symmetry assignment of lowest wavenumber mode at 145 cm$^{-1}$ was not conclusive ($A_g$, $B_g$ or the superposition of both symmetries was proposed) [96,97] (Table 1 gather the different assignment proposed in the literature). Here, thanks to the different pressure dependences, we confirm that, at ambient conditions, one soft mode and one hard mode with different symmetries are superimposed at 145 cm$^{-1}$. At pressure above $P_c$= 13.9(1) GPa, the soft mode changes its behavior and starts hardening, which marks the isostructural $M_1$-$M_1$' transition. This transition is reversible with no pressure hysteresis. Extrapolating the $v_{SM}^2(P)$ to $v_{SM}$=0 limit gives $P_c^*$= 26.9(4) GPa, for the potential stability limit of the $M_1$ phase. The ratio between the slopes d$v^2$/dP below and above $P_c$ is 2.4(1) close to 2, characteristic of a continuous phase transition. With increasing pressure, the hardening mode successively cross the $A_{g(1)}$ mode at 25 GPa, and the 190 cm$^{-1}$ $A_{g(2)}$ mode at 29.5 GPa, and shows a deviation from the linear dependence at pressure higher than 32 GPa before disappearing at 41 GPa (Figure 7(a)). The spectra recorded between 20 and 29 GPa showing the successive crossing between low wavenumber modes are reported in figure S6(a). The pressure evolution of the half width at half maximum (HWHM) of both $B_{g(1)}$ and $A_{g(1)}$ modes obtained from the decomposition of the Raman spectra using Fityk software are reported in figure S6(b). The $B_{g(1)}$ HWHM is narrower (2 cm$^{-1}$) than the $A_{g(1)}$ (6 cm$^{-1}$). Under pressure, the HWHM of $B_{g(1)}$ remains constant as $A_{g(1)}$ decreases sharply. Above 20 GPa, depending on experience and therefore local conditions, HWHM may fluctuate, but as far as positions are concerned, everything is reproducible. The integrated intensity (area) progressively increases with pressure above $P_c$ (Figure 7(b)). Contrary to previous studies [66,67,68,71,73], the $A_{g(2)}$ mode at 190 cm$^{-1}$ does not show any abrupt increase is the rate dv/dP at $P_c$. We rather measured a small decrease of the slope from dv/dP= 0.36(1) cm$^{-1}$/GPa to 0.22(1) cm$^{-1}$/GPa at the transition. The discontinuity reported at 10 GPa in previous studies might be a consequence of the use of non-hydrostatic pressure transmitting media, i.e. NaCl, KCl [66,67], or ethanol-methanol [68,69,80] that are known to be strongly anisotropic at this pressure. The half width at half maximum, HWHM, (Figure 7(d)) shows a regular decrease with pressure up to 29 GPa, followed by a tendency to increase that is always observed at such high pressure because of the progressive loss of hydrostaticity of the helium transmitting medium. The same tendency is measured on the ruby pressure marker (see Figure 7(d)). The integrated intensity (area) of the $A_{g(2)}$ peak (Figure 7(c)) is almost constant up to 32 GPa and suddenly drops at higher pressure before disappearing at 41 GPa. The pressure dependences dv/dP and the Grüneisen parameters of the Raman modes are reported in Table 2 and their positions are given at 0 GPa for the $M_1$ phase and at 13.9 GPa for the $M_1'$ high-pressure phase.

A second original observation in Raman spectra of the $M_1$ phase under pressure is the splitting of the mode at 225 cm$^{-1}$ in two components at pressure as low as 3 GPa within the resolution limit of our spectrometer (see Figure 7(a) and figure S7). This mode was in the past associated to a single $A_g$ symmetry but experimental [97] and theoretical studies [98,99] have proposed that two modes of $A_g$ and $B_g$ symmetries could be superimposed at room condition. Here again, pressure allows for distinguishing both modes due to their different pressure-dependences. Both modes show a sharp slope changes in v(P) at $P_c$= 13.9(1) GPa (see Table 2).

The variations of the spectral features at the transition allows for correlating the Raman modes with the different components of the strain. The spontaneous shift v($M_1'$) - v($M_1$) is calculated after subtracting the wavenumber v($M_1$) extrapolated above $P_c$ from the behavior measured below 14 GPa. The $A_{g(3)}$ scales linearly with the absolute value $|e_{22}|$ of the spontaneous strain along $b_{M1}$ (Figure 8(a)) or with ($e_{33}$-$e_{22}$) that reflects the deformation of the ($b_{M1}$,$c_{M1}$) plane. The $B_{g(3)}$ scales linearly with the square of the spontaneous strain along $c_{M1}$ ($e_{33}^2$)(see Figure 8(b)). With further increasing pressure, at 29 GPa, another discontinuity is observed in the splitting (see figure S7). In previous studies, the splitting was observed, only above 27-28 GPa [71,69] or above 19 GPa [73] but was associated to phase X or to a new insulating $M_3$ phase, different from phase X.

A third original observation in the $M_1$ phase, concerns the Raman modes $B_{g(2)}$ at 260 cm$^{-1}$, and $A_{g(4)}$ at 310 cm$^{-1}$. They exhibit an unusual small pressure-dependence of their positions (see Figures 9(a) and 9(b)). The slopes are dv/dP = 0.03(1) cm$^{-1}$/GPa and dv/dP = 0.13(1) cm$^{-1}$/GPa, respectively (see Table 2). However, they show an abrupt change in their dv/dP at $P_c$. To the best of our knowledge, the $B_{g(2)}$ slope discontinuity was never reported. Some authors have seen that this mode disappear between 14-15 GPa [71,69] or at 22 GPa [74]. From our observations, the intensity starts decreasing at 14 GPa but the mode is still observed up to 30 GPa. The slope changes of the $A_{g(4)}$ mode was reported at the $M_1$-$M_1'$ transition above 13 GPa [71,74]. The HWHM (not shown) exhibits a regular decrease with increasing pressure similar to that measured on the $A_{g(2)}$ mode (Figure 7(b)). The spontaneous shift v($M_1'$) - v($M_1$) for the $A_{g(4)}$ mode scales linearly with $e_{33}^2$ (Figure 9(c)).

The Raman modes at higher wavenumbers exhibit classical increase of their positions with increasing pressure (see figure S8). The slopes dv/dP are larger than those measured for the low-wavenumber modes. A small decrease of the slopes dv/dP is observed at $P_c$ (see Table 2). Notice that the slope of the $B_{g(4)}$ mode, at 340 cm$^{-1}$ and $B_{g(8)}$ mode, at 665 cm$^{-1}$ are almost not affected by the transition at $P_c$.

At 32 GPa, the collapse of the Raman intensity and the sudden increase of the background are the signature of the formation of the metallic phase X. With further increasing pressure up to 41 GPa, the Raman peaks disappear and some new peaks appear progressively. The Raman signature of the pure phase X recorded during decompression is reported in Figure 10 (in red at 28.7 GPa) and shows nine weak peaks at 185, 325, 440, 466, 505, 662, 707, 763 and 845 cm$^{-1}$ (see Fig. 10 and Fig. S8). Upon decompression, the Raman spectra show a transformation, between 22 and 18.5 GPa, to a spectrum of reasonable intensity that is not compatible with neither the $M_1'$ nor $M_1$ structures but can be explained by a coexistence between phase X and a new structure. The coexistence persists down to 9.3 GPa but, between 5 and 3 GPa, phase X completely disappeared and the remaining spectrum resembles that of the triclinic T phase (or $M_3$) measured on 0.7% Cr-doped $VO_2$ by Marini at al. [67]. The same signature was reported on $VO_2$ nanoparticles below 23.9 GPa and down to 2.1 GPa by Li at al. [74], and was interpreted as a back transformation from the baddeleyite-type $M_X$ phase into a new baddeleyite-type $M_X'$ phase with a local structure similar to the $M_1$ structure.

## 4. **Discussion**

### *4.1. First transition from $M_1$ to $M_1'$ at 14 GPa*

Depending on the pressure transmitting medium, the $M_1$-$M_1'$ transition has been reported at pressure varying between 10 and 15 GPa [66,67,68,72,71,74,80]. In our hydrostatic conditions, $VO_2$ single crystal exhibit a first isostructural transition, $M_1$ to $M_1'$, at $P_c$= 13.9(1) GPa as observed by Raman and x-ray diffraction measurements. The transition is quasi-continuous, second order-like with no measurable volume jump. The transition is displacive with oxygen displacements compatible with the R-point condensation (in the parent rutile) without strong modification of the VV dimers nor of the twist angle of vanadium chains (Fig. 4). The oxygen sub-lattice spontaneous displacements and the spontaneous deformation of the ($b_{M1}$, $c_{M1}$) plane follow the same quadratic dependence with pressure (Fig. 2 and 4). The monoclinic $a_{M1}$ lattice parameter is not affected by the transition (Fig. 1).

We can combine these new high-quality experimental data with reliable information published so far, and suggest therefore a coherent picture of phase transitions in $VO_2$ compressed and heated/cooled. The rutile to monoclinic transition is an *improper ferroelastic transition of displacive type* and is induced by the four-component order-parameter spanning $R_1^-$ irreducible representation at the R-point of the tetragonal Brillouin zone [36,37,10,11,38,39,47]. Mechanical (vibrational) representation of the rutile-type structure at the R-point of the Brillouin zone reads:

$$T_M = (3R_1^-)_V + (3R_1^- + 3R_1^+)_O \tag{1}$$

Thus, the symmetry-breaking atomistic mechanism of the structural R-$M_1$ transformation contains simultaneous vanadium and oxygen atoms displacements, both transforming as $R_1^-$ and, therefore, coupled bilinearly in the free-energy. In other words, the symmetry lowering and distortion of the tetragonal structure are controlled by coupled vanadium and oxygen displacements. In the high-temperature rutile phase, the four components of the $R_1^-$ OP are zero: $\eta_1 = \eta_2 = \eta_3 = \eta_4 = 0$, and the vanadium chains are regularly aligned with fixed VV bonds distances of 2.86 Å. At the rutile to $M_1$ transition, one component of the $R_1^-$ OP takes non-zero value ($\eta_1 \neq 0, \eta_2 = \eta_3 = \eta_4 = 0$). The R-point imposes that two antiferroelectric vanadium displacements occurs when $\eta_1 \neq 0$: one along the $a_{M1}$ axis ($a_{M1}=2c_R$) forming VV dimers on one chain and one off-axis in the plane perpendicular to $a_{M1}$ axis forming twisted vanadium on the nearest neighbor vanadium chain [10,47]. Thus if $\eta_1 \neq 0$ two twisted vanadium chains with VV dimers are formed in the $M_1$ phase.

Under pressure, the second component $\eta_2$ of the $R_1^-$ OP, which reduces its symmetry to $B_{g(1)}$ after the Brillouin zone folding, drives the structure transformation to the $M_2$ phase with: $\eta_1 = \eta_2 \neq 0$ ($\eta_3 = \eta_4 = 0$). One set of vanadium chains pair (VV dimers) is not twisted while the other set stays twisted but loses the VV dimers. The $M_2$ phase is expected at 27 GPa, as estimated from extrapolating the linear part of the soft mode wavenumber $v_{SM}^2(P)$ measured experimentally to the $v_{SM}$=0 limit. However, the $M_2$ phase is not observed because the isostructural $M_1$-$M_1'$ transition occurs at $P_c$= 13.9(1) GPa suppressing this instability and conserving the $M_1$-type phase thermodynamically more stable. The observed phonon softening does not drive the $M_1$-$M_1'$ transition (but drives the $M_1$-$M_2$). The isostructural transition somehow prevents the $M_1$-$M_2$ transition from taking place under hydrostatic pressure as detailed in the Landau-based analysis developed in the next section. The oxygen displacements and the monoclinic ($b_{M1}$, $c_{M1}$) plane distortion remind those found in the rutile to $CaCl_2$ transition observed in $VO_2$ at higher temperature [71,79,80] and in many other $AO_2$ oxides. However, the oxygen polyhedron is not only rotating along to the $a_{M1}$ axis (former $c_R$ axis in the rutile phase). Let us show, in the framework of phenomenological theory, that the reason for the isostructural transition lays in the highly anharmonic dependence of the free-energy on the non-totally-symmetric OP.

### 4.2. Understanding the $VO_2$ phase diagram from phenomenological theory

Our experimental findings allow us to derive a complete picture of phase transitions in $VO_2$ observed under different pressure and temperature conditions (P<35 GPa). This requires to consider two-component effective order-parameter. The image-group, reduced form of the relevant four-dimensional representation $R_1^-$ to a two-dimensional effective order-parameter group, possesses the point-symmetry 4mm. Phenomenological models for the two-dimensional tetragonal image-group were analysed in detail by Y. Gufan and co-workers and cited in [100].

The basic invariants forming the integrity basis for the image-group 4mm are:

$$I_1 = \eta_1^2 + \eta_2^2, \text{ and } I_2 = \eta_1^2 \cdot \eta_2^2. \quad (2)$$

Accordingly, the most compact structurally stable order-parameter ten-degree expansion, which is necessary to account for two consecutive first-order phase transitions R-$M_1$ and $M_1$-$M_1$' (see Appendix 1), is expressed as:

$$F(\eta_1, \eta_2, P, T) = a_1(P,T)I_1 + a_2(P,T)I_1^2 + b_1 I_2 + c_{12} I_1 I_2 + a_4 I_1^4 + b_2 I_2^2 + a_5 I_1^5. \quad (3)$$

The free-energy (3) has four minima corresponding to the four phases known for $VO_2$:

I: $\eta_1 = \eta_2 = 0$ ~ R;
II: $\eta_1 \neq 0, \eta_2 = 0$ ~ $M_1$;
III: $\eta_1 = \eta_2 \neq 0$ ~ $M_2$;
IV: $\eta_1 \neq \eta_2 \neq 0$ ~ T.

Figure 11 shows a section of the theoretical phase diagram, corresponding the potential (3), which is topologically adequate to understand the $VO_2$ pressure-temperature phases diagram experimentally mapped in hydrostatic conditions. In addition, this free-energy expansion includes the existence of a <u>critical end-point</u> K (gas-liquid type) on the $M_1$-$M_1$' transition line at which the first-order transition transforms to a cross-over continuous regime (see annex 1). The fact that no apparent volume could be experimentally measured indicates that the isostructural $M_1$-$M_1$' transition is quasi-continuous and reveals that the pressure path passes in close vicinity of this critical point K (Figure 11). Varying pressure at higher temperature should allow to measure an increasing volume jump at the $M_1$-$M_1$' transition as one moves away from the critical point. The topology of the phenomenological phase diagram also predicts that the triclinic T structure can be observed at even higher hydrostatic pressures. On the contrary, $M_2$ and rutile R phases might hardly be formed under hydrostatic pressure at ambient temperature. The phase diagram explains also that Cr-doped $VO_2$, which adopt triclinic (for 0.7% Cr) or $M_2$ (for 2.5% Cr) phases, are reported to first transform to $M_1$ phase at 2.7 GPa and 3.7 GPa, respectively, and then, to the same $M_1$' phase as pure $VO_2$ at 12 GPa [67,72].

It is worth stressing that the general form of Eq. (3) and the diagram shown in Figure 11 are generic ones since they also account for stress/strain effects. Indeed, we can distinguish two types of strain components: (i) through improper spontaneous strains induced by the primary order-parameter $R_1^-$ ($e_{11}$, $e_{22}$, $e_{33}$, $e_{12}$, and $e_{13}$), and (ii) through external deviatoric stress ($e_{23}$) developing under quasi-hydrostatic compression conditions, or surface effects in thin films, for instance [41,42,43]. Although the coupling terms in the free-energy have different forms, $\eta_i^2 \cdot e_{jk}$ and $\eta_i^2 \cdot e_{lm}^2$, they should be integrated with the unique quadratic invariant $I_1$ in the free-energy (3). This will lead to renormalizing the corresponding coefficient ($a_1+c_{ij}+c_{il}$)→$\tilde{a}_1$ but without modifying the general form of Eq. (3). The topology of the phase diagram of Fig.11 remain unchanged, however the transition line can be shifted and then the $M_2$ phase could be observed under non-hydrostatic stress. Topology means the correct description of the phases in contact, and prediction of the order for the phase transition that can occur between them.

The changes in the midinfrared transmittance/reflectance [66,67,68,69] and in the resistivity observed previously under pressure [71,70,80] are concomitant with the $M_1$-$M_1$' isostructural transition. This strongly suggests that electronic properties and structural modifications (with oxygen displacements) are linked and that the Peierls mechanism is valid. We can assume that this isostructural transition can also be induced by uniaxial/bi-axial stresses in thin films, or in non-stoichiometric $VO_2$ for which internal stresses can be generated. Thus, experimental studies that have questioned the Peierls mechanism because of the observation of a monoclinic-like metallic $VO_2$ where electronic and structural transitions seem decoupled [101,102,103,104] did not consider the possibility of having formed the isostructural $M_1$' phase.

### 4.3. Raman signature of $M_1$, $M_2$ or T phases as a tool for thin film engineering

The technological interest on $VO_2$ has led to the study of various thin films or nanobeams using Raman spectrometry as a valuable tool to differentiate between rutile, $M_1$, $M_2$ or T phases [105,55,45,57,106,107,108,45,97,30,109]. The metallic rutile has a weak signal composed of broad modes at 300 and 550 cm$^{-1}$ (for $A_{1g}+B_{1g}+E_g$) [110] that are difficult to measure. The Raman signature of $M_1$ is quite well documented but not all the 9$A_g$ and 9$B_g$ modes were observed and the symmetry assignments are still being debated (see Table 1). The present study, thanks to pressure-induced variations of the peak positions allows for clarifying the assignment (see section 3.2). Moreover, very few is known on the atomic displacements (eigenvectors) involved in each mode. Since the Raman study under oxygen isotopic substitution [67], it is often said that the two intense low wavenumber modes at 190 and 225 cm$^{-1}$ involve predominantly vanadium displacements. This was supported by the phonon density of state obtained with *ab initio* calculations [111,98,112,99,6,113]. There is a widespread belief that these modes are associated with the stretching and twisting features of the dimerized chains and contribute to the $M_1$-rutile transition [17,67,111,98,69,99]. However, these modes do not obviously soften at the MIT [92,13,94]. The Raman signature of the T phase is similar to that of $M_1$ but the $A_g(1)$ mode is downshifted to 126 cm$^{-1}$, the $A_g(2)$ is upshifted to 200 cm$^{-1}$, and a small splitting of the $A_g(3)+B_g(3)$ is observed [67,45,106,114,115,116,117]. In the $M_2$ phase, the $A_g(1)$ mode downshifts even more to 50 cm$^{-1}$, the $A_g(2)$ stays at 200 cm$^{-1}$, and two components are clearly observed for the $A_g(3)+B_g(3)$ [67,105,55,45,57,106,108,116]. We do not endorse the fact that the $A_g(1)$ mode could be a breathing mode of spin-Peierls dimerized 1-D spin ½ Heisenberg chain [116] but rather found that the two modes $A_g(1)+B_g(1)$ at 145 cm$^{-1}$ are the vanadium displacive modes expected from the condensation of the Rutile $R_1^-$ OP. The progressive softening of the $A_g(1)$ mode through $M_1$ to T and $M_2$ structural transformation, where one half of the the Peierls pairing and twisting are partially removed or with increasing pressure, where only one mode softens until the transition to $M_1$' hinders this instability, supports our finding.

The splitting of the $A_g(3)+B_g(3)$ mode, at 225 cm$^{-1}$, in both $M_1$ and $M_1$' phases, highlighted in Figures 7(a) and Figure S7, was often misunderstood in the past. Several DFT calculations concluded that the zigzag V motions that untwist the VV pairs are located between ?6.0 THz (197 cm$^{-1}$) [98], 6.38 THz (213 cm$^{-1}$) [111] or ≈6.5 THz (217 cm$^{-1}$) [118], close to the positions of the A$_g$(3)+B$_g$(3) modes. We do not observe any softening with pressure and doubt that these modes are linked to the pairing or tilting motions of VV dimers. We found that the splitting is observed in the M$_1$ phase at 2-3 GPa (Figure 7(a)). Indeed, from the linear pressure evolution of each mode, we found that the two modes intersect at 1.9 GPa confirming they have different symmetries. The angular dependence of the Raman intensity in different polarized conditions measured outside the DAC as shown in Figure S9, also show the superimposition of two different symmetries already in the M$_1$ monoclinic phase at ambient conditions in agreement with Shibuya at al. [97]. The splitting is equal to 0.8 cm$^{-1}$ in the M$_1$ at room condition which explains why it was hardly detectable in past studies. At the M$_1$-M$_1$' iso-structural transition, both modes display an abrupt change of their dv/dP (see Figures 7(a) and S7 or in Table 2). We found that the A$_g$(3) scales linearly with the spontaneous strain along $b_{M1}$ whereas the B$_g$(3) scales linearly with the spontaneous strain along $c_{M1}$ (see Figure 8(a-b)). Thus, the A$_g$(3)+B$_g$(3) splitting is a good marker of the nature of the strain experienced by VO$_2$. The unusual pressure behaviour observed at 24-25 GPa (see Figure 7(a) and S7) is a consequence of the saturation of the spontaneous deformation along $b_{M1}$ axis while the one along the $c_{M1}$ increases without there being a phase transition. Quantification of the monoclinic deformation can also be done using either the A$_g$(4) mode at 310 cm$^{-1}$ or the B$_g$(2) mode at 260 cm$^{-1}$. In the M$_1$ stability region, below P$_c$, both modes are insensitive to hydrostatic compression (see Figs. 9(a) and 9(b)) and accurate wavenumbers measurements beyond the possible drifts of the equipment can be done using the B$_g$(2) mode as an internal reference. Above P$_c$, the A$_g$(4) scales linearly with $e_{33}^2$ (see Fig. 9(c)) or $e_{Total}^2$ (not shown) whereas the B$_g$(2) scales linearly with $e_{33}^4$ or $e_{Total}^4$.

The high-wavenumbers modes, B$_g$(4) at 340 cm$^{-1}$ (see Fig. S11) or B$_g$(8) at 665 cm$^{-1}$ (not shown), scale linearly with the monoclinic volume (M$_1$ or M$_1$') with no measurable discontinuity at P$_c$= 13.9(1) GPa. The A$_g$(9) mode at 615 cm$^{-1}$ and Ag(5)+Bg(5) doublet at 389/393 cm$^{-1}$ scale linearly with the octahedron volume (see Fig. S11(c-d)). The apparent discontinuity in the v(P) at P$_c$ is due to the non-linear pressure dependence of the oxygen octahedron volume (see Fig. 3c).

**Conclusion**

The phase diagram of VO$_2$ has been investigated in the past but several aspects remained unclear. Indeed, the influence of non-hydrostatic components induced either by the pressure-transmitting medium or the form of the sample (powder vs single crystal) on the phase transition led to some discrepancies. Here, we present a combined X-ray diffraction and Raman spectroscopy investigations of high-quality VO$_2$ single crystal under pressure using Helium as the pressure transmitting medium. For the first time, a pressure-induced soft mode is observed. This behaviour is supposed to drive a transition towards a M$_2$ phase at pressure around 26 GPa. However, an intermediate phase transition is observed at 13.9 GPa, hindering this phonon instability. The isostructural nature of the phase transition at 13.9 GPa is confirmed experimentally. The microscopic mechanism is clarified and is based on the displacements of oxygen atoms. A phenomenological analysis based on the Landau theory of phase transition is proposed to describe the P-T phase diagram. Considering a strong anharmonic potential, the phase transitions, including the isostructural one, are described. The coupling with strains can explained the shift of the transition lines found in doped VO$_2$ or in thin films. At higher pressure, a phase transition to a metallic phase, probably triclinic, is observed starting from 32-35 GPa. On decompression, this phase transforms to another triclinic structure. Using high-pressure allows for separating overlapping peaks at ambient conditions and brings some new insights into the assignment of the different modes observed in Raman spectra. In addition, the results of the Raman spectroscopy allow relating some vibrational to different strain components or to pressure-induced microscopic variations such as the octahedron volume. This opens the opportunity to characterize the thin films in terms of structure, nature and amplitude of strain.

**Acknowledgements**

PB acknowledges the French CNRS for financial support through Tremplin@INP2020 and C. Goujon, C. Felix, Ch. Bouchard, A. Prat and J. Debray (CNRS, Institut Néel Grenoble) and J. Jacobs (ESRF Grenoble) for their technical help. The authors are grateful to ESRF ID15b for in-house beamtime allocation. Ch. Lepoittevin (UGA Institut Néel Grenoble) and W.A. Crichton (ESRF Grenoble) are also acknowledged for their advice on crystallographic questions. LB acknowledges the PROCOP Mobilität Programm for financing two months visit in Grenoble. LN2 is a joint International Research Laboratory (IRL 3463) funded and co-operated in Canada by Université de Sherbrooke (UdS) and in France by CNRS as well as ECL, INSA Lyon, and Université Grenoble Alpes (UGA). It is also supported by the Fonds de Recherche du Québec Nature et Technologie (FRQNT).

**Annex 1**

Although the active order parameter $R_1^-$ is four-components [119,120], only one single component is relevant to account for the rutile to $M_1$ structure distortion and becomes non-zero in the low-symmetry phase. This allows considering for $R_1^-$-$M_1$ an effective phenomenological model with one-component order parameter. The $R_1^-$ symmetry forbids odd-degree terms in a free-energy expansion, and we get a canonical form for the Landau potential expanded to the tenth-degree:

$$F(\eta,P,T) = a_1(P,T)\,\eta^2 + a_2(P,T)\,\eta^4 + a_3\eta^6 + a_4\eta^8 + a_5\eta^{10}. \tag{A1}$$

The mathematical analysis of the model has been performed by Gufan [120] who concluded that the minimal degree of $F(\eta)$ required to describe two consecutive first-order phase transitions (here, R-$M_1$ and $M_1$-$M_1'$) is ten (see also [100] and references therein). This model allows to describe two low-symmetry phases. These phases have identical symmetries but differ by the magnitude of the order-parameter $\eta$. Therefore, the isostructural transition is intrinsically included into this description. Figure A1 shows the evolution of the theoretical phase diagram with increasing power of the free-energy $F$. Thus, for $VO_2$ undergoing two discontinuous phase transitions, R-$M_1$ and $M_1$-$M_1'$ (Fig.A1(c)), the phenomenological model (A1) is sufficient assuming $a_4<0$ and $a_5>0$, with $M_1$-$M_1'$ phase transition being iso-structural. Notice that to choose the maximal degree of expansion (between eight and ten) the main point is the character (continuous or discontinuous) of the first transition R-$M_1$.

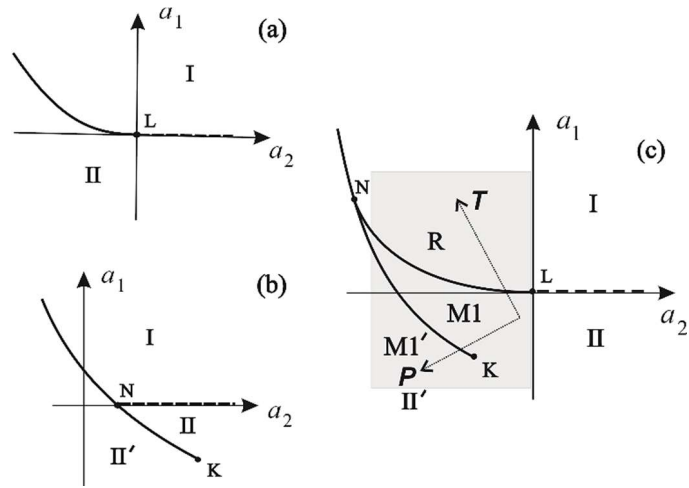

Figure A1. Equilibrium phase diagram corresponding to the free-energy (A1) in the plane of phenomenological coefficients ($a_1,a_2$) for: (a) canonical six-degree expansion ($a_3>0$, $a_4=a_5=0$); (b) eight-degree potential ($a_3<0$, $a_4>0$, $a_5=0$); (c) ten-degree expansion ($a_3<0$, $a_4<0$, $a_5>0$). Figure (c) schematically shows "pressure ($P$)-temperature ($T$)" plane (grey area, dotted axes). Solid line – first-order, dashed – second-order phase transition lines. L – Landau tricritical point, N – triple point, K – critical end-point of the iso-structural phase transition.

**Figure 1:**

(color online) VO$_2$ monoclinic cell parameters with increasing pressure; (a) $a_{M1}$ axis, (b) $b_{M1}$ axis, (c) β angle between $a_{M1}$ and $c_{M1}$, and (d) $c_{M1}$ axis in the M$_1$ phase $P2_1/n$ cell choice 2. The full lines correspond to BM EoS between 0 and 34 GPa (see text). Insert in figure (a) shows the VV dimmers along the monoclinic $a_{M1}$ axis. Insert in figure (b) shows the VO6 octahedra in the ($b_{M1}$, $c_{M1}$) plane.

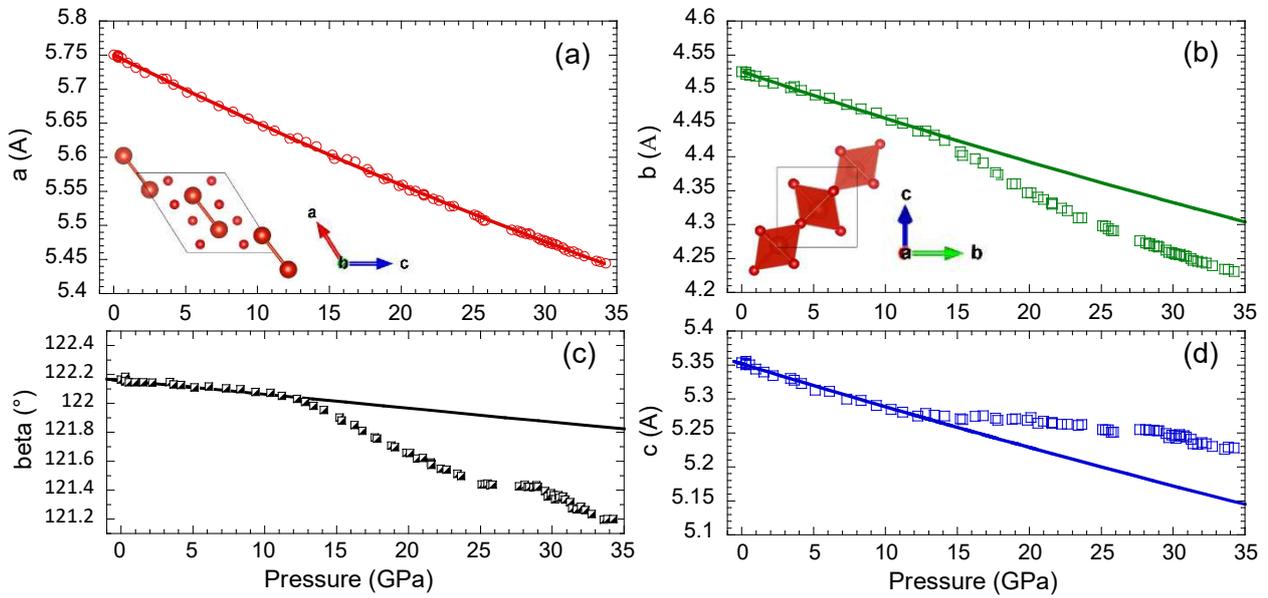

**Figure 2:**
(color online) Spontaneous deformations calculated in the high-pressure monoclinic cell against the original low pressure monoclinic $M_1$ using the EoS extrapolated above 14 GPa. The full lines are square root functions with (P-$P_c$) with fixed $P_c$=13.9 GPa.

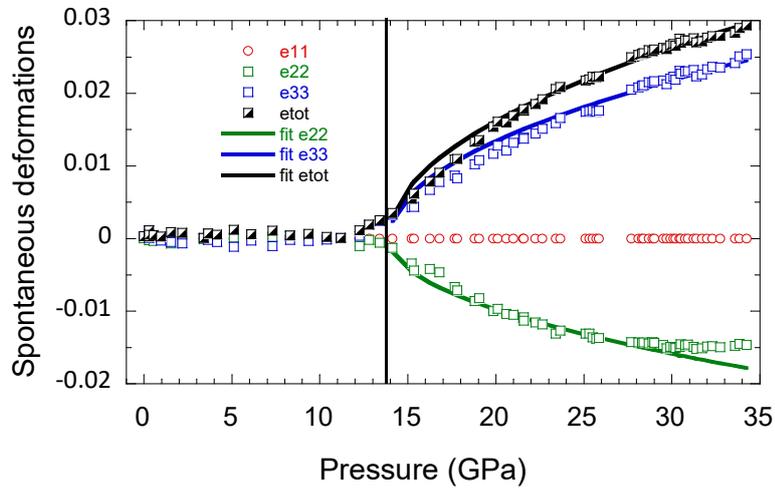

**Figure 3:**

(color online) VO$_2$ monoclinic parameters with increasing pressure (a) volume for Z=4 and F-f plot in the insert, (b) vanadium distances inside VV dimmers, (c) volume of VO$_6$ polyhedron. The full lines correspond to third order BM EoS. Others EoS from previous works are reported for comparison.

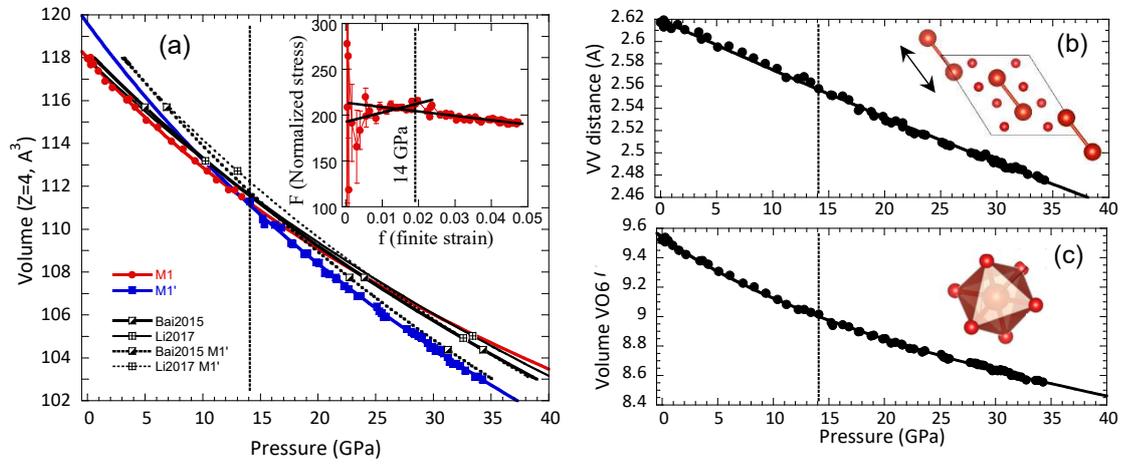

**Figure 4:**
(color online) Pressure dependence of (a) Vanadium fractional coordinates V(x)+0.712, V(y)-0.478 and V(z)+0.473 and (b) fractional oxygen displacements measured after subtracting the displacements extrapolated from the behavior below 14GPa in space-group *P2₁/n* (n°14, Z=4, cell choice 2). The full lines in (b) are square root function with (P- $P_c$) with fixed $P_c$=13.9 GPa.

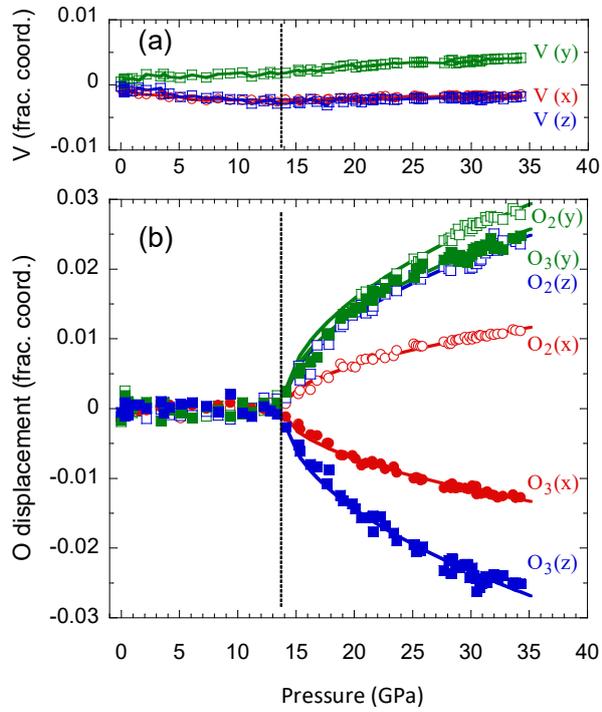

**Figure 5:**

(color online) Result of the LeBail profile fitting in the triclinic P-1 unit-cell at 35 GPa for phase X. Expected diffraction peaks are indicated by ticks. The difference between the experimental and the fit is reported at the bottom. Insert show the two-dimensional image of the crystal with tentative indexation.

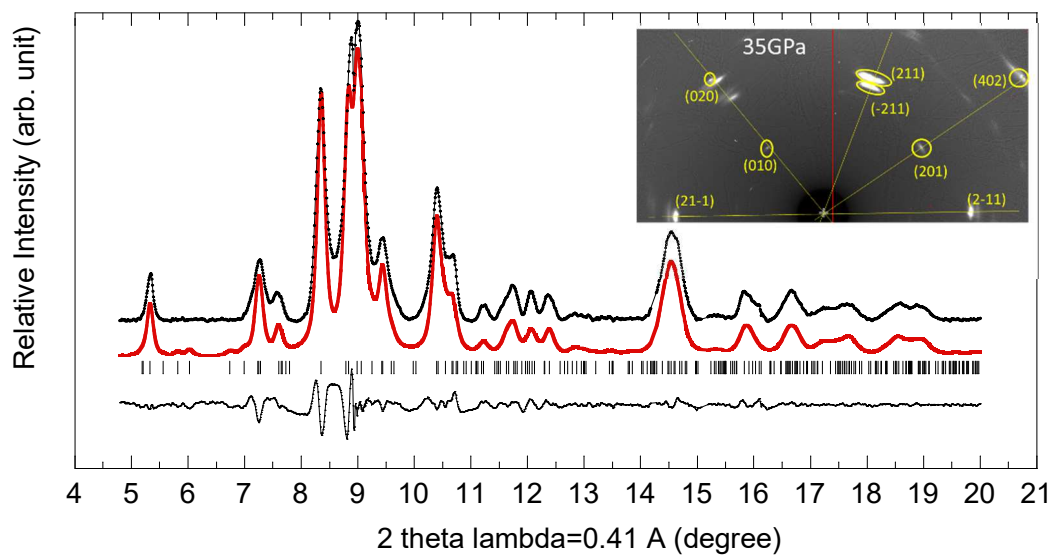

**Figure 6:**
(color online) Low wavenumber part (70-340cm$^{-1}$) of the Raman spectra measured on VO$_2$ single crystal showing the softening/hardening of the 145 cm$^{-1}$ Raman mode under increasing pressure. Pressures are quoted on the left of each spectrum. Black lines correspond to monoclinic M$_1$ phase. Red lines highlight the pressure higher than Pc= 13.9(1) GPa. The symmetry A$_g$ or B$_g$ of each mode is indicated at the bottom.

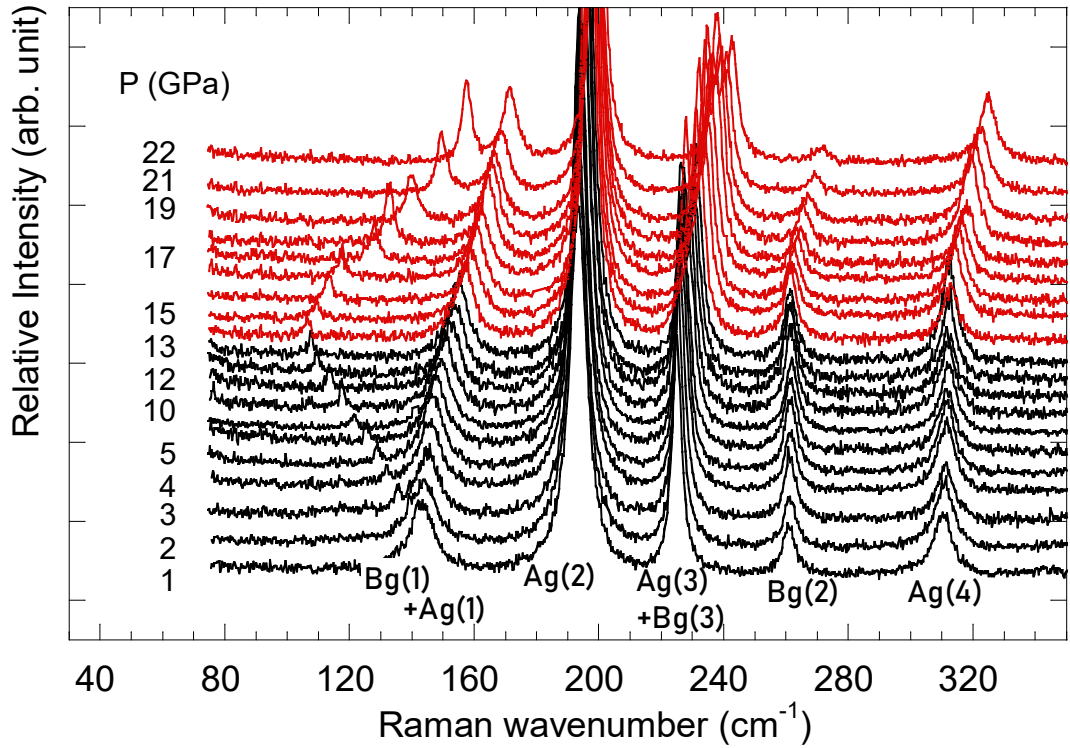

**Figure 7:**

(color online) Low wavenumber Raman spectral parameters measured under increasing hydrostatic pressure; (a) wavenumbers of the Ag(1), Bg(1), Ag(2), Ag(3) and Bg(3) modes (see labels in figure 6), (b) integrated intensity (area/s) of the Ag(1) and Bg(1) modes, (c) integrated intensity (area/s) of the most intense Ag(2) mode, (d) HWHM of the Ag(2) mode and of the Ruby pressure marker. Red and blue colours stand for Ag and Bg symmetry respectively.

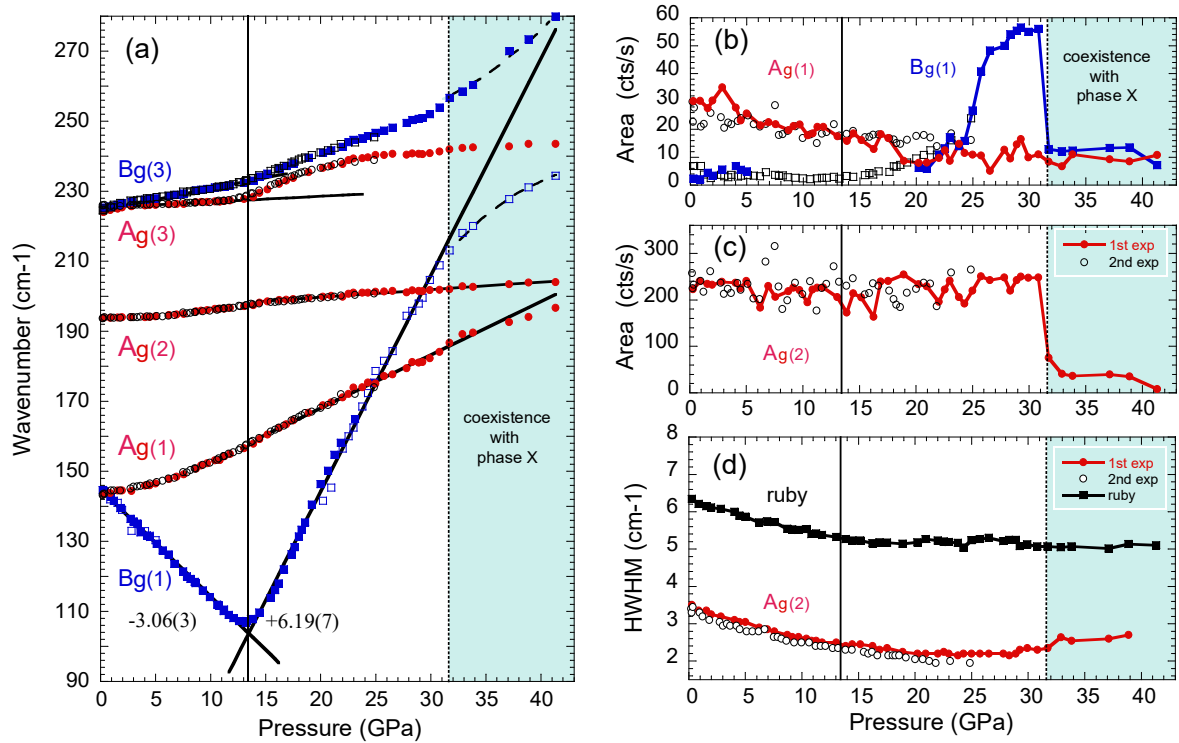

**Figure 8:**

(color online) Spontaneous Raman shift w(M$_1$')-w(M$_1$) after subtracting the wavenumber w(M$_1$) extrapolated above P$_c$ from the behavior before P$_c$=14 GPa (a) Ag(3) and Bg(3) at 225 cm$^{-1}$ against |e$_{22}$| spontaneous strain along $b_{M1}$ and (c) Bg(3) at 225 cm$^{-1}$ against e$_{33}$$^2$ spontaneous strain along $c_{M1}$.

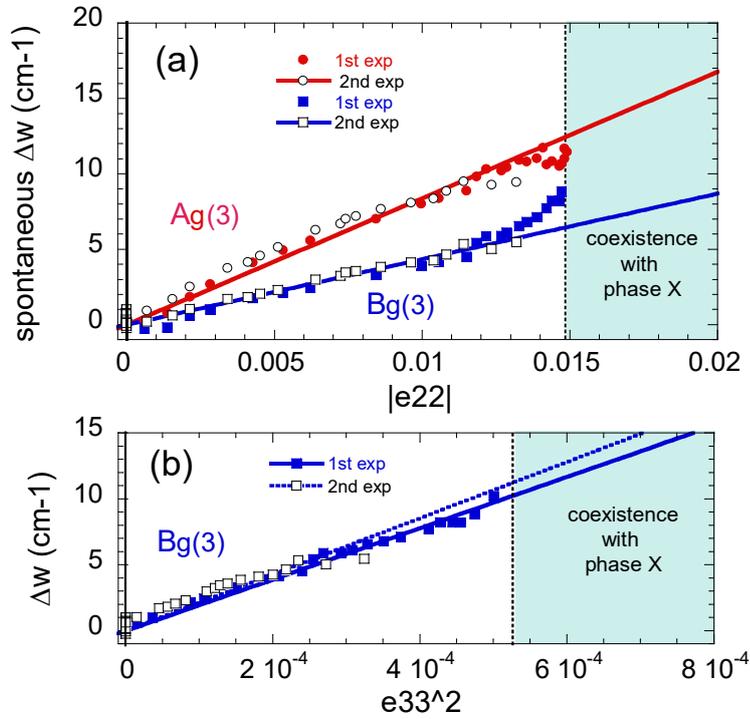

**Figure 9:**
(color online) Wavenumber of the Raman modes measured under increasing hydrostatic pressure on $VO_2$. (a) $B_{g(2)}$ at 260 cm$^{-1}$, and (b) $A_{g(4)}$ at 310 cm$^{-1}$. The slopes $d\nu/dP$ are reported in different pressure regions. (c) $A_{g(4)}$ at 310 cm$^{-1}$ spontaneous Raman shift against $e_{33}^2$ spontaneous strain along $c_{M1}$.

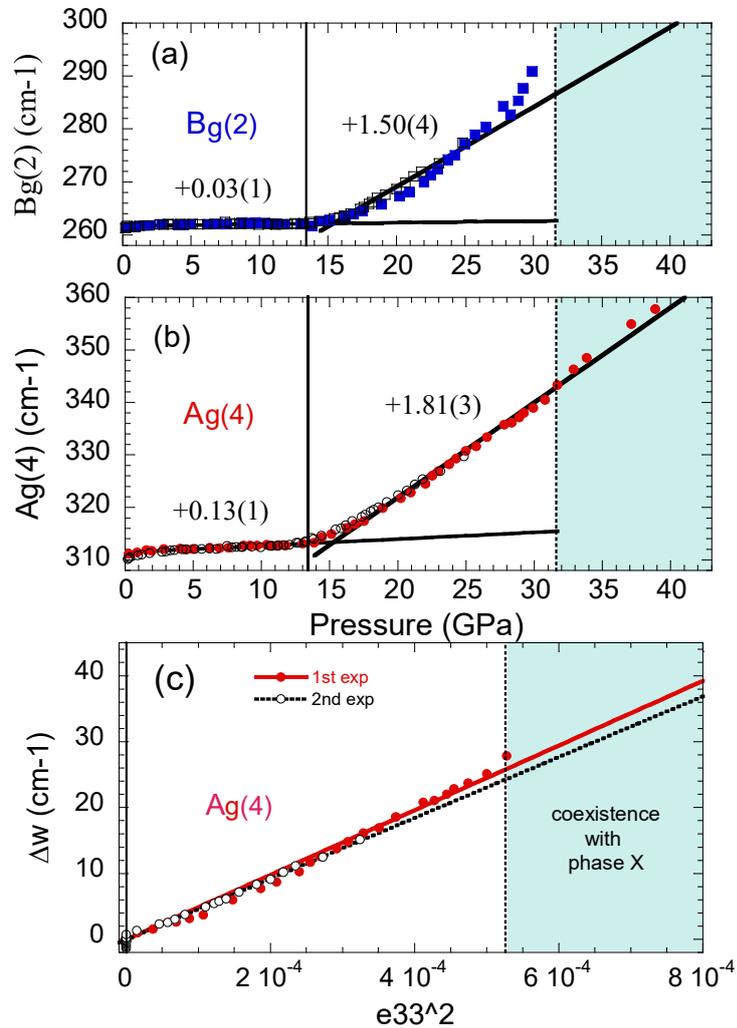

**Figure 10:**

(color online) Raman spectrum measured during decompression from 42 GPa. The signature of Phase X (in red at 28.7 GPa) is maintained up to 22 GPa and is transformed to a triclinic phase between 22 and 18.5 GPa. A coexistence between both structures is observed up to 3 GPa. A strained triclinic phase is retained at atmospheric pressure and room temperature. The spectra a corrected from a linear background.

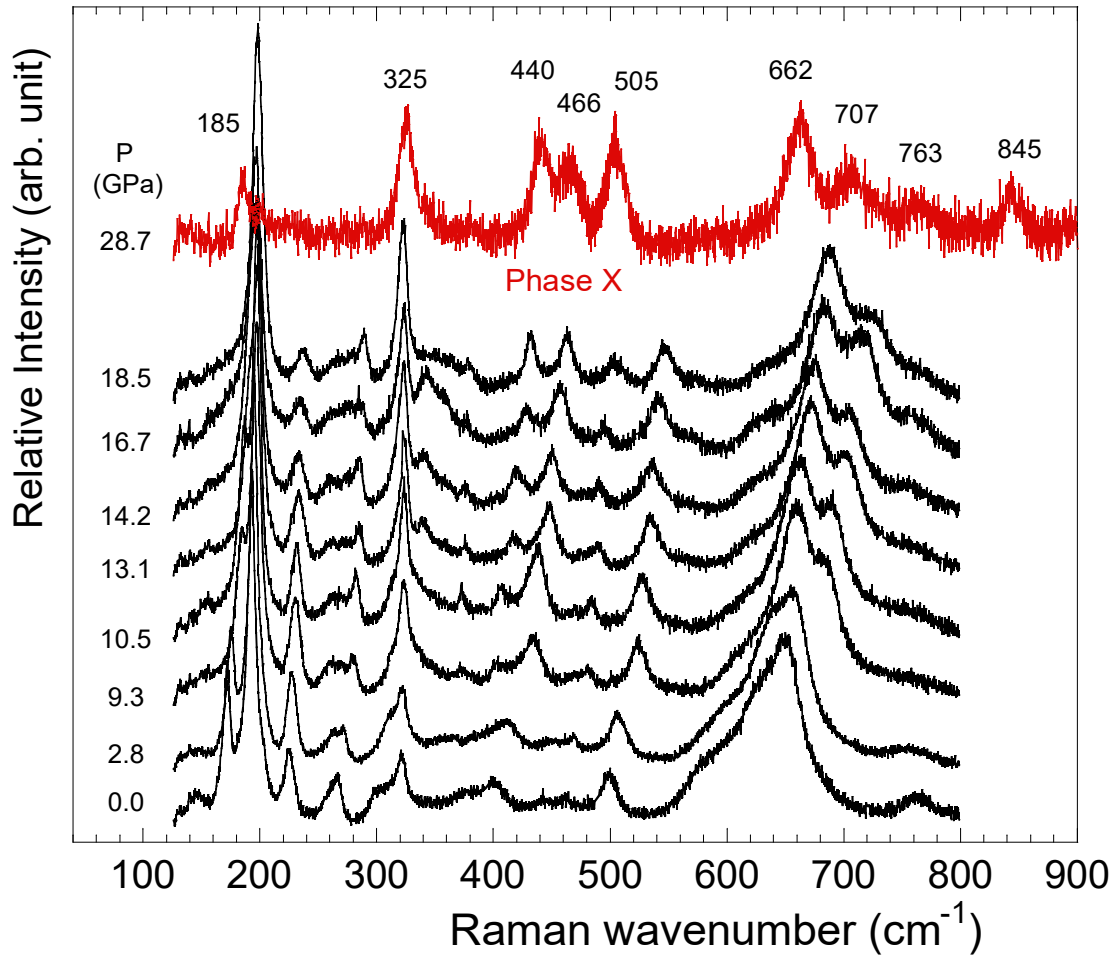

**Figure 11:**
Equilibrium phase diagram corresponding to the free-energy (3) in the plane of phenomenological coefficients $(a_1, a_2)$ for $0 < a_2 < (c_{12}^2/4b_2)$, $c_{12} < 0$, $a_4 < 8b_2$, $a_5 > 0$. Solid line – first-order, dashed – second-order phase transition lines. K – critical end-point, $N_1$ and $N_2$ are three-phase points. Pressure (P) and temperature (T) axes are shown schematically.

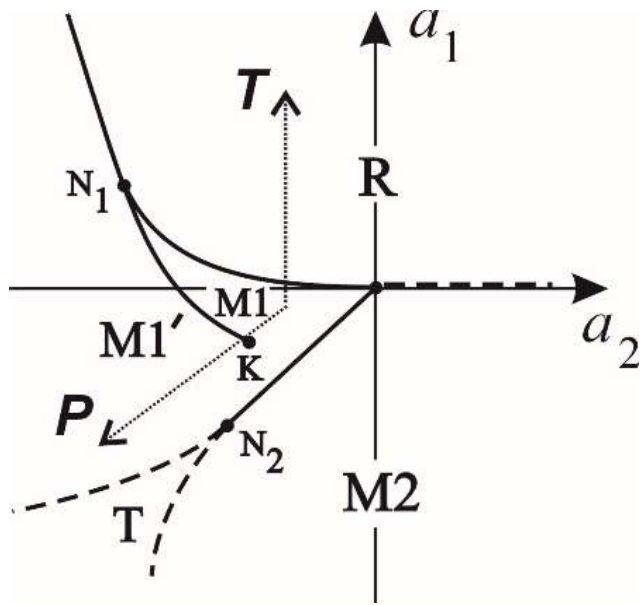

**Figure A1:**
Equilibrium phase diagram corresponding to the free-energy (2) in the plane of phenomenological coefficients ($a_1, a_2$) for: (a) canonical six-degree expansion ($a_3 > 0$, $a_4 = a_5 = 0$); (b) eight-degree potential ($a_3 < 0$, $a_4 > 0$, $a_5 = 0$); (c) ten-degree expansion ($a_3 < 0$, $a_4 < 0$, $a_5 > 0$). Figure (c) schematically shows "pressure (P)-temperature (T)" plane (grey area, dotted axes). Solid line – first-order, dashed – second-order phase transition lines. L – Landau tricritical point, N – triple point, K – critical end-point of the iso-structural phase transition.

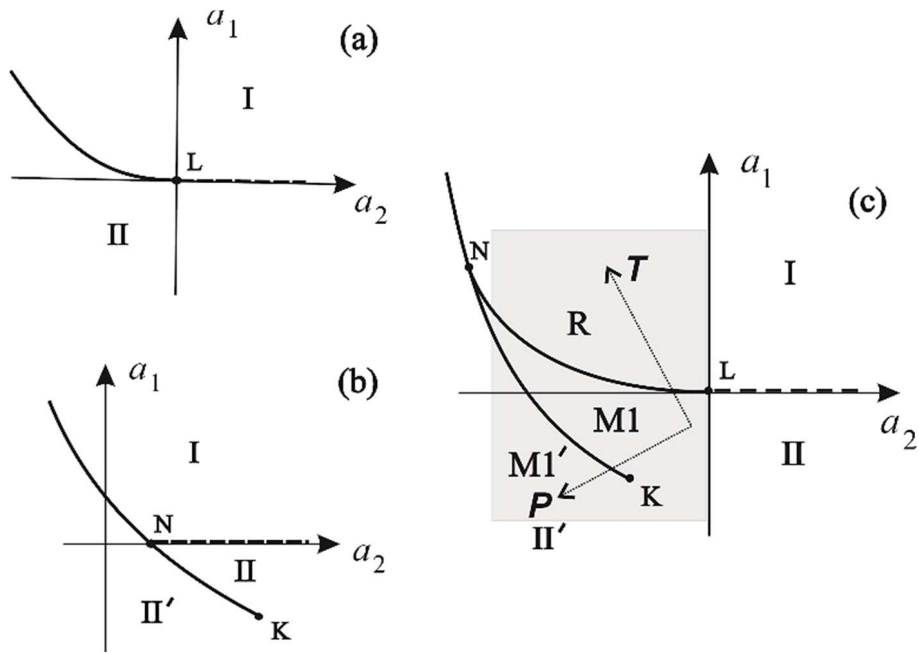

**Table 1.**
Position in wavenumber (cm$^{-1}$) of the Raman active modes measured experimentally or calculated for monoclinic $M_1$ phase of $VO_2$ and symmetry assignment propositions from literature. Lines in dark gray or light gray highlight the Ag or Bg symmetry of the Raman modes. The stars ***, **, * indicate the Raman intensity from most intense to less intense.

| Position (cm$^{-1}$) | | [92] | [93] | [94] | [17] | [95] | [98] calc. | [96] | [97] | [99] calc. | Monoclinic $M_1$ This work |
|---|---|---|---|---|---|---|---|---|---|---|---|
| ** | 145 | - | - | $A_g$ | $B_g$ | - | $A_g$ | $B_g$ | $A_g+B_g$ | $A_g$ | $A_g(1)+B_g(1)$ |
| *** | 190 | $A_g$ | $A_g$ | $A_g$ | $A_g$ | $A_g$ | $A_g+B_g$ | $A_g$ | $A_g$ | $A_g+B_g$ | $A_g(2)$ |
| *** | 225 | $A_g$ | $A_g$ | $A_g$ | $A_g$ | $A_g$ | $A_g+B_g$ | $A_g$ | $A_g+B_g$ | $A_g+B_g$ | $A_g(3)+B_g(3)$ |
| * | 260 | $A_g$ | $B_g+B_g$ | $B_g+B_g$ | - | - | $B_g+B_g$ | $B_g$ | $B_g$ | $B_g$ | $B_g(2)$ |
| ** | 310 | $A_g$ | $A_g$ | $A_g$ | - | $B_g$ | $A_g$ | $A_g$ | $A_g$ | $A_g$ | $A_g(4)$ |
| | 340 | $A_g$ | $B_g$ | $B_g$ | - | $A_g$ | $A_g$ | $A_g$ | $A_g$ | $A_g+B_g$ | $B_g(4)$ |
| ** | 390 | $A_g$ | $A_g$ | $A_g$ | - | $A_g$ | $A_g$ | $A_g$ | $A_g$ | $A_g$ | $A_g(5)$ |
| ** | 394 | - | $B_g$ | $B_g$ | - | - | $B_g$ | $B_g$ | $B_g$ | - | $B_g(5)$ |
| | 440 | $A_g$ | $B_g$ | $B_g$ | - | - | $B_g$ | $B_g$ | $B_g$ | $B_g$ | $A_g(6)$ |
| | 445 | - | $B_g$ | $A_g$ | - | - | $B_g$ | $B_g$ | - | $B_g$ | $B_g(6)$ |
| | 485 | - | $B_g$ | $B_g$ | - | - | $B_g$ | $B_g$ | $B_g$ | $B_g$ | $B_g(7)$ |
| | 500 | $A_g$ | $A_g$ | $A_g$ | - | $A_g$ | $A_g$ | $A_g$ | $A_g$ | $A_g$ | $A_g(7)$ |
| ** | 595 | $B_g$ | $A_g$ | $A_g$ | - | - | $B_g$ | $A_g$ | $B_g$ | $B_g$ | $A_g(8)$ |
| *** | 615 | $A_g$ | $A_g$ | $A_g$ | - | $A_g$ | $A_g$ | $A_g$ | $A_g$ | $A_g$ | $A_g(9)$ |
| | 665 | - | $B_g$ | $B_g$ | - | - | $A_g$ | $B_g$ | $A_g$ | $A_g$ | $B_g(8)$ |
| | 850 | - | $B_g$ | $B_g$ | - | - | $B_g$ | $B_g$ | $B_g$ | $B_g$ | $B_g(9)$ |

**Table 2:**
(color online) Wavenumber dependence with pressure for Raman modes in $M_1$ and $M_1'$ high pressure monoclinic $VO_2$. SM= Soft mode. Bg modes in green. Grüneisen parameter $\Upsilon=(K/\nu).(d\nu/dP)_T$. Errors are in parenthesis.

| Raman mode Symmetry | Position @00GPa (cm$^{-1}$) | Slope (cm$^{-1}$/GPa) | Grüneisen $\Upsilon$ | Position @13.9GPa (cm$^{-1}$) | Slope (cm$^{-1}$/GPa) |
|---|---|---|---|---|---|
| Ag(1) | 142.9(2) | +0.77(4) | +1.04(9) | 158.9(3) | +1.52(2) |
| Bg(1) | 144.9(2) | -3.06(3) | -4.1(2) | 106.4(7) | +6.19(7) |
| Ag(2) | 192.5(1) | +0.36(1) | +0.36(2) | 198.5(1) | +0.22(1) |
| Bg(3) | 224.6(2) | +0.62(2) | +0.53(4) | 233.6(1) | +1.15(1) |
| Ag(3) | 225.4(1) | +0.16(1) | +0.14(1) | 229.3(2) | +1.10(4) |
| Bg(2) | 261.7(1) | +0.03(1) | +0.022(8) | 259.9(2) | +1.50(4) |
| Ag(4) | 311.4(1) | +0.13(1) | +0.081(9) | 311(4) | +1.81(3) |
| Bg(4) | 340.7(3) | +4.42(4) | +2.52(11) | 400.1(4) | +4.06(4) |
| Ag(5) | 388.8(3) | +4.06(4) | +2.03(9) | 444.1(3) | +2.79(3) |
| Bg(5) | 392.8(5) | +4.32(6) | +2.13(10) | 450.7(3) | +2.61(3) |
| Bg(6) | 442 | ---- | ---- | ---- | ---- |
| Ag(6) | 442.8(3) | +2.35(4) | +1.03(5) | 477(1) | +3.15(9) |
| Bg(7) | 483 | ---- | ---- | ---- | ---- |
| Ag(7) | 499.3(1) | +2.64(1) | +1.03(4) | 536.2(2) | +2.02(2) |
| Ag(8) | 594.5(8) | +4.37(11) | +1.43(8) | 654.5(6) | +2.78(12) |
| Ag(9) | 613.4(2) | +3.86(2) | +1.22(5) | 668.8(3) | +2.37(3) |
| Bg(8) | 662.8(7) | +2.85(9) | +0.83(5) | 703.8 | +2.72(7) |
| Bg(9) | ---- | ---- | ---- | ---- | ---- |